\documentclass[letterpaper,twocolumn,prx,aps,superscriptaddress,floatfix,longbibliography]{revtex4-2}
\usepackage[latin1]{inputenc}
\usepackage[T1]{fontenc}
\usepackage{bm}
\usepackage{amssymb,amsbsy,amsmath}
\usepackage{graphicx}
\usepackage{verbatim}
\usepackage{color}
\usepackage{pifont}
\usepackage{hyperref}
\usepackage[dvipsnames]{xcolor}
\usepackage{verbatim}
\usepackage{esvect}
\usepackage{float}
\usepackage{algorithm,algpseudocode,float}
\usepackage{lipsum}

\makeatletter

\makeatletter

\renewenvironment{widetext@grid}{%
  \par\ignorespaces
  \setbox\widetext@top\vbox{%
   \vskip15\p@
   \hb@xt@\hsize{%
    \leaders\hrule\hfil
    \vrule\@height6\p@
   }%
   \vskip6\p@
  }%
  \setbox\widetext@bot\hb@xt@\hsize{%
    \vrule\@depth6\p@
    \leaders\hrule\hfil
  }%
  \onecolumngrid
  \let\set@footnotewidth\set@footnotewidth@ii
}{%
  \par
  \twocolumngrid\global\@ignoretrue
  \@endpetrue
}%

\makeatother

\hypersetup{
	colorlinks=true,       		% false: boxed links; true: colored links
	linkcolor=orange,          	% color of internal links
	citecolor=blue,             % color of links to bibliography
	urlcolor=blue,          % color of external links
}
\usepackage{braket}

\begin{document}

\title{A hybrid-frequency on-chip programmable synthetic-dimension simulator with arbitrary couplings}
    \affiliation{Laboratory of Quantum Information, University of Science and Technology of China, Hefei 230026, China}
    \affiliation{Anhui Province Key Laboratory of Quantum Network, University of Science and Technology of China, Hefei 230026, China}
    \affiliation{CAS Center For Excellence in Quantum Information and Quantum Physics, University of Science and Technology of China, Hefei 230026, China}
    \affiliation{Hefei National Laboratory, University of Science and Technology of China, Hefei 230088, China}
    \affiliation{Quantum Science Center of Guangdong-Hong Kong-Macao Greater Bay Area, Shenzhen 518045, China}
    
    \author{Xiao-Dong~Zeng}
    \thanks{These authors contributed equally to this work.}
    \affiliation{Laboratory of Quantum Information, University of Science and Technology of China, Hefei 230026, China}
    \affiliation{Anhui Province Key Laboratory of Quantum Network, University of Science and Technology of China, Hefei 230026, China}
    \affiliation{CAS Center For Excellence in Quantum Information and Quantum Physics, University of Science and Technology of China, Hefei 230026, China}
    \author{Zhao-An Wang}
    \thanks{These authors contributed equally to this work.}
    \affiliation{Laboratory of Quantum Information, University of Science and Technology of China, Hefei 230026, China}
    \affiliation{Anhui Province Key Laboratory of Quantum Network, University of Science and Technology of China, Hefei 230026, China}
    \affiliation{CAS Center For Excellence in Quantum Information and Quantum Physics, University of Science and Technology of China, Hefei 230026, China}
    \affiliation{Quantum Science Center of Guangdong-Hong Kong-Macao Greater Bay Area, Shenzhen 518045, China}
    \author{Jia-Ming~Ren}
    \thanks{These authors contributed equally to this work.}
    \author{Yi-Tao Wang}
    \email{yitao@ustc.edu.cn}
    \author{Chun~Ao}
    \author{Wei~Liu}
    \affiliation{Laboratory of Quantum Information, University of Science and Technology of China, Hefei 230026, China}
    \affiliation{Anhui Province Key Laboratory of Quantum Network, University of Science and Technology of China, Hefei 230026, China}
    \affiliation{CAS Center For Excellence in Quantum Information and Quantum Physics, University of Science and Technology of China, Hefei 230026, China}
    \author{Nai-Jie~Guo}
    \affiliation{Laboratory of Quantum Information, University of Science and Technology of China, Hefei 230026, China}
    \affiliation{Anhui Province Key Laboratory of Quantum Network, University of Science and Technology of China, Hefei 230026, China}
    \affiliation{CAS Center For Excellence in Quantum Information and Quantum Physics, University of Science and Technology of China, Hefei 230026, China}
    \affiliation{Hefei National Laboratory, University of Science and Technology of China, Hefei 230088, China}
    \author{Lin-Ke~Xie}
    \affiliation{Laboratory of Quantum Information, University of Science and Technology of China, Hefei 230026, China}
    \affiliation{Anhui Province Key Laboratory of Quantum Network, University of Science and Technology of China, Hefei 230026, China}
    \affiliation{CAS Center For Excellence in Quantum Information and Quantum Physics, University of Science and Technology of China, Hefei 230026, China}
    \author{Jun-You~Liu}
    \affiliation{Laboratory of Quantum Information, University of Science and Technology of China, Hefei 230026, China}
    \affiliation{Anhui Province Key Laboratory of Quantum Network, University of Science and Technology of China, Hefei 230026, China}
    \affiliation{CAS Center For Excellence in Quantum Information and Quantum Physics, University of Science and Technology of China, Hefei 230026, China}
    \affiliation{Hefei National Laboratory, University of Science and Technology of China, Hefei 230088, China}
    \author{Yu-Hang~Ma}
    \author{Ya-Qi~Wu}
     \affiliation{Laboratory of Quantum Information, University of Science and Technology of China, Hefei 230026, China}
    \affiliation{Anhui Province Key Laboratory of Quantum Network, University of Science and Technology of China, Hefei 230026, China}
    \affiliation{CAS Center For Excellence in Quantum Information and Quantum Physics, University of Science and Technology of China, Hefei 230026, China}
    \author{Shuang~Wang}
    \affiliation{Laboratory of Quantum Information, University of Science and Technology of China, Hefei 230026, China}
    \affiliation{Anhui Province Key Laboratory of Quantum Network, University of Science and Technology of China, Hefei 230026, China}
    \affiliation{CAS Center For Excellence in Quantum Information and Quantum Physics, University of Science and Technology of China, Hefei 230026, China}
    \affiliation{Hefei National Laboratory, University of Science and Technology of China, Hefei 230088, China}
    \author{Pei-Yun~Li}
    \affiliation{Laboratory of Quantum Information, University of Science and Technology of China, Hefei 230026, China}
    \affiliation{Anhui Province Key Laboratory of Quantum Network, University of Science and Technology of China, Hefei 230026, China}
    \affiliation{CAS Center For Excellence in Quantum Information and Quantum Physics, University of Science and Technology of China, Hefei 230026, China}
     \author{Zong-Quan~Zhou}
   \affiliation{Laboratory of Quantum Information, University of Science and Technology of China, Hefei 230026, China}
    \affiliation{Anhui Province Key Laboratory of Quantum Network, University of Science and Technology of China, Hefei 230026, China}
    \affiliation{CAS Center For Excellence in Quantum Information and Quantum Physics, University of Science and Technology of China, Hefei 230026, China}
    \affiliation{Hefei National Laboratory, University of Science and Technology of China, Hefei 230088, China}
       \author{Mu~Yang}
     \affiliation{Laboratory of Quantum Information, University of Science and Technology of China, Hefei 230026, China}
    \affiliation{Anhui Province Key Laboratory of Quantum Network, University of Science and Technology of China, Hefei 230026, China}
    \affiliation{CAS Center For Excellence in Quantum Information and Quantum Physics, University of Science and Technology of China, Hefei 230026, China}
    \author{Jin-Shi~Xu}
   \author{Xi-Wang Luo}
     \author{Jian-Shun~Tang}
    \email{tjs@ustc.edu.cn}
    \author{Chuan-Feng~Li}
    \email{cfli@ustc.edu.cn}
    \author{Guang-Can~Guo}
    \affiliation{Laboratory of Quantum Information, University of Science and Technology of China, Hefei 230026, China}
    \affiliation{Anhui Province Key Laboratory of Quantum Network, University of Science and Technology of China, Hefei 230026, China}
    \affiliation{CAS Center For Excellence in Quantum Information and Quantum Physics, University of Science and Technology of China, Hefei 230026, China}
    \affiliation{Hefei National Laboratory, University of Science and Technology of China, Hefei 230088, China}

	\date{\today}
	
	\renewcommand{\figurename}{Fig.}
	
	\newcommand{\Todos}[1]{\textcolor{red}{#1}}
			
\begin{abstract}
%The synthetic frequency dimension in photonics has emerged as a potentially powerful paradigm for simulating high-dimensional physical systems through low-dimensional devices. Driven by recent advancements in thin-film lithium niobate platforms, this provides a highly programmable and scalable foundation for large-scale on-chip photonic simulators. To realize practical on-chip synthetic frequency lattice simulators, the extension to higher dimensions and the introduction of long-range coupling and asymmetric coupling are crucial, which imposes stringent experimental requirements. Here we propose the method that combines intra-resonant lattice points and inter-resonant modes, demonstrating arbitrarily coupled lattice models. Using our devices, we experimentally realize well-known models such as the Hall ladder, Creutz ladder, and Su-Schrieffer-Heeger model, including their long-range coupled forms. We directly observe the band structures in the quasi-momentum space and important phenomena such as spin-momentum locking, topological flat band, and Aharonov-Bohm cage effect. Additionally, we propose that cascading our devices could enable applications in piecewise-continuous optical frequency shifting. Our results demonstrate the powerful potential for simulating complex lattices---with both symmetric and asymmetric couplings---offering promising insights for future large scale on-chip simulation devices.

High-performance photonic chips provide a powerful platform for analog computing, enabling the simulation of high-dimensional physical systems using low-dimensional devices with additional synthetic dimensions. The realization of large-scale complex simulations necessitates an architecture capable of arbitrary coupling configurations (encompassing symmetric, asymmetric and long-range coupling schemes) which is also crucial for scaling up. Previous approaches rely on excessive physical components to introduce asymmetric coupling, however, are restricted in reconfiguring and scaling by  the relatively complicated structures \cite{sridhar2024,li2023light,yu2021topological}. Here, to solve this problem, we propose a hybrid-frequency synthetic-dimension simulator architecture that combines both intra-resonant and inter-resonant frequency-lattice sites, and experimentally demonstrate it using the thin-film lithium niobate (TFLN) photonic chip. Employing this hybrid programmable architecture, we are able to simulate both the regular and long-range coupled forms of diverse compound-lattice models, such as the Hall ladder, Creutz ladder (symmetric) and Su-Schrieffer-Heeger (SSH, asymmetric) model, on a single chip, simultaneously reducing the experimental requirements significantly. As results, the direct readout of the bandstructure of the SSH model is able to be achieved, to be distinguished from all previous works, and important phenomena such as spin-momentum locking, topological flat band and Aharonov-Bohm cage effect are also observed with lower experimental requirements. Furthermore, applications like piecewise-continuous optical frequency shifting can be enabled by cascading our devices. Our results offer promising insights for future large-scale complex on-chip simulators with arbitrary couplings.

\end{abstract}

\maketitle

\noindent\large\textbf{Introduction}

\begin{figure*}[t]
    \centering
    \includegraphics[width=1\textwidth]{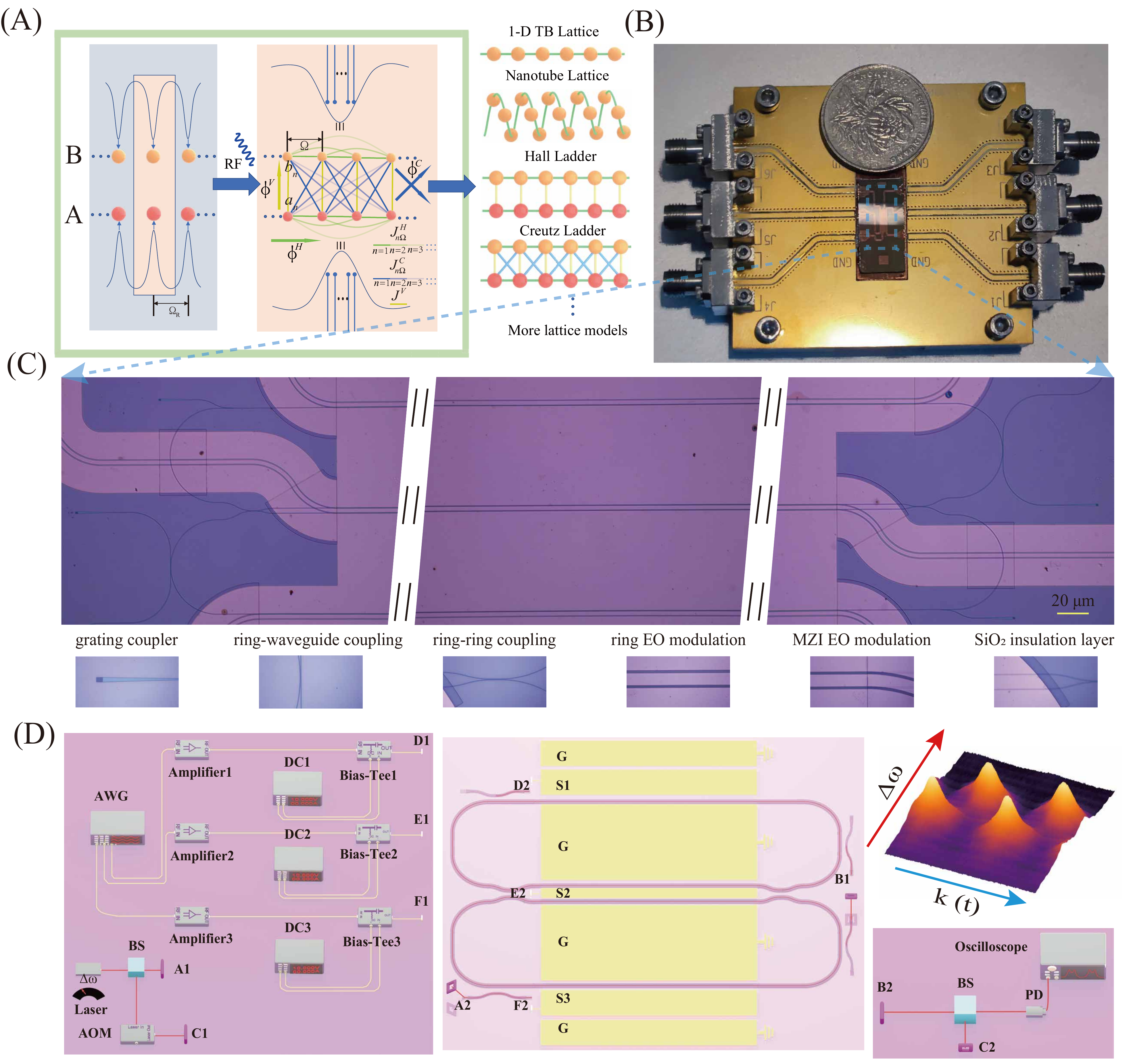}
    \caption{\textbf{Scheme and experimental setup.} (\textbf{A}) The dynamics of light with resonator under periodic electro-optic modulation. Each resonant peak of the resonator, separated by the FSR $\Omega_R$, hosts a series of frequency-lattice sites selected by low-frequency RF signal $\Omega$. By coupling the intra-resonant frequency sites of two resonators, complex lattice models can be engineered. The lattice network incorporates three types of couplings: local modulation on the resonator (green), along with DC (yellow) and RF (blue) signals applied to the MZI. (\textbf{B}) Image of the thin film lithium niobate (TFLN) photonic chip. (\textbf{C}) The optical microscope images of the device consisting of LN racetrack microresonator and electrodes. The bottom insets showcase the key devices fabricated on TFLN. (\textbf{D}) Schematic of setup for detecting band structures and photon distribution on sites% with a MZI-assisted race-track resonators on TFLN
    . Three DC+RF signal sets are applied via Bias-Tees to ground-signal-ground (GSG) electrodes, controlling resonators ($S_1$, $S_3$) and MZI ($S_2$). The RF signal is generated by a microwave source or an arbitrary waveform generator (AWG), and is connected to a power amplifier before connected to the Bias-Tees. The letter pairs (A1,A2)$\sim$(F1,F2) denote transmission connections between the panels for electrical circuits and optical paths. Band structure detection employs a $\Delta\omega$-detuned probe laser, with time-resolved transmittance (revealing quasi-momentum slices) captured by a photodetector (PD) and oscilloscope. The top-right figure presents a characteristic band structure reconstructed from temporally stacked transmittance spectra. The acousto-optic modulator (AOM) is used in the heterodyne detection optical path to shift the frequency of a portion of the incident laser, and the site distribution measurements utilize the heterodyne optical path with a PD.}%coupling among lattice points of each lattice due to local modulation on each resonator (orange); nearest neighbouring coupling between lattices due to the DC signals on MZIs (blue); cross (long-range neighbouring) coupling between lattices due to the RF signals on MZIs (green).
    \label{fig:scheme}
\end{figure*}

\normalsize
%模拟的量子模拟介绍、光学合成频率维度介绍
 Analog quantum simulators, which use programmable and scalable devices to emulate Hamiltonians of various physical phenomena, offer a promising path to realize the advantages of quantum computing before large-scale digital quantum computers become feasible \cite{Altman2021,Daley2022,Cirac2012,dai2024,Shao2024}. By constructing periodic systems in a synthetic rather than real space, researchers can investigate higher-dimensional physical phenomena within lower-dimensional platforms, and effectively simulate complex physical behaviors in more accessible systems \cite{Fang2012,chen2021,lustig2019,ozawa2019,TopoRev,sridhar2024,Luo2015,luo2017,regensburger2012}. In photonics, a widely adopted approach to constructing synthetic dimensions is through the frequency domain, in which optical frequency modes are mapped to coupled sites to perform analog simulation \cite{Yuan18,Yuan2019,Yuan2021}. Early advances in this field were largely driven by fiber-based resonator architectures \cite{Dutt2019,Dutt2020,Dutt2022,Li2021,WangK2021,wang2021nature}. Notably, the realization of synthetic spaces on chip-scale platforms offers considerable advantages compared to table-top systems, such as better programmability, enhanced scalability, and greater operational stability, which are challenging for table-top systems.

%On-chip的光学频率维度模拟的发展，基于FSR模式的发展，其实验难度和扩展性低
 Recently, thin-film lithium niobate (TFLN) has emerged as a promising optical signal manipulation platform \cite{hu2025,zhu2021,feng2024,knaut2024}. Its low-loss, high-quality photonic integration together with strong Pockels effect enables superior modulation performance, making it an excellent platform for photonic integrated circuits and future photonic interconnects \cite{LNRev1,LNRev2,zhang2019,Zhang2017,yu2022,xu2020,zhu2024twenty}. The strong Pockels effect also makes TFLN particularly suitable for frequency-domain simulation. Based on the TFLN platform and leveraging the synthetic frequency dimension, implementations have already been achieved for phenomena such as quantum random walks, mirror-induced reflection, multilevel Rabi oscillations, and tight-binding models, among others \cite{Javid2023,Hu2020,Dutt2022,Dinh2024,saxena2023}. Considering the size of the chip, the free spectral range (FSR) of the resonator on the TFLN chip is typically on the order of 2$\pi \times$10 GHz \cite{Ye-2025,liu2025}. Applying such high-frequency modulation signals, particularly multiharmonic signals, is challenging because the high cost of the required equipment. In our previous work \cite{wza2024}, on the contrary, we proposed using a modulation frequency much lower than the resonator linewidth to construct sites within a resonant peak (intra-resonant frequency sites) \cite{Hu2020,Martin2017}. This significantly unlocks the available dimension used to develop the scalability of on-chip photonic simulators.

%各种拓扑模型的重要性和引入，阐述在合成频率维度中怎么实现这些模型
%拓扑光子学、非厄米光子学、平带物理（讲很多格点模型需要交叉项、长程耦合很重要、非均匀晶格很重要）
 %With the development of the field, it is imperative to simulate and explore more complex and richer lattice models by synthetic frequency dimensions. %
 Achieving long-range coupling between sites is essential for expanding the dimensionality and interaction complexity within synthetic spaces \cite{wang2020light,wang2021nature}. Notably, implementing long-range coupling among intra-resonant sites substantially reduces experimental constraints compared to conventional approaches for on-chip applications \cite{wza2024}. Moreover, lattice structures with nonuniform connectivities can exhibit richer physics in real space \cite{li2023light,yu2021topological,qiao2023topological}. The implementation is challenging with only intra-resonant frequency sites, while is straightforward by adding another resonator (chain) equipped with asymmetric cross couplings. Throughout this paper, when the forward/backward hopping from one chain to the other (intra-resonant sites) is conjugate, we term it as symmetric coupling, otherwise we term it as asymmetric coupling (still Hermitian). Extending intra-resonant sites to higher dimensions and combining them with inter-resonant frequency sites can enable richer lattice models (including both symmetric and asymmetric couplings), leading to more powerful quantum simulators.

%引入腔内模式+FSR模式的混合晶格模拟，可以实现原本只靠FSR模式调制不能实现的晶格
In this work, we develop a fully-programmable, highly-scalable photonic hybrid-frequency-mode synthetic dimension simulator on a TFLN chip by combining both intra-resonant and inter-resonant sites. Using a single chip with this optimized synthetic dimension beyond conventional schemes, we are able to simulate richly featured lattice models  \cite{SSH1979}. %In our previous work, we created lattice points using intra-resonant peak and verified that directly observing band structures could be performed through time-resolved transmission spectroscopy. Compared to conventional methods that rely on adjacent mode coupling, the requirements for radio-frequency (RF) modulation are reduced by three orders of magnitude, while still demonstrating equally rich simulation results \cite{wza2024}. 

By applying RF modulation far below the resonant mode linewidth (which induces the inter-resonant coupling) on both ring resonators and Mach-Zender interferometer (MZI), we realize Hall and Creutz ladder lattices (symmetric coupling) with distinct magnetic fluxes \cite{Dutt2020,liu2025,ye2025observing,Hung2021,creutz1999,wang2025versatile}, observing spin-momentum locking, topological flat bands and the Aharonov-Bohm (AB) cage effect \cite{wang2025versatile,Vidal1998,Vidal2001,He2021}. Furthermore, we propose that by leveraging the AB cage effect and cascading the chips, it is possible to achieve piecewise-continuous optical frequency shifting applications \cite{li2025weakly,Hu2021shift,Balla2023,Eggleton2019,wright2017,wang2025versatile}. Moreover, due to the flexibility of lattice construction using intra-resonant sites, we can easily introduce long-range coupling to realize coupled Hall ladder and coupled Creutz ladder. 

Most importantly, by combining intra-resonant and inter-resonant sites, we construct nonuniform-connectivity lattices in long-range scheme. The engineering of lattices with asymmetric couplings enables realization of richer physical models. As an example, we construct the classic asymmetric coupled lattice model---the Su-Schrieffer-Heeger (SSH) model (one of the simplest topological models, represented by 1D dimer chains), and directly observe its band structures which to our knowledge has not yet been realized in photonic systems. 

We demonstrate the high scalability and programmability of intra-resonant sites, and showcase the advantages of hybrid-frequency sites in constructing lattices with nonuniform connectivities, providing new insights into the simulation of large-scale topological lattice models and non-Abelian lattice gauge fields \cite{cheng2025non,sridhar2024,lin2018}.

\noindent\large\textbf{Theoretical framework}

\normalsize
By utilizing an MZI-assisted device and performing appropriate detuning and modulations, we can realize arbitrarily-coupled lattice models \cite{wang2025versatile}, including symmetric and asymmetric couplings. For symmetrically-coupled models, since the forward and backward hopping are conjugate, the off-diagonal terms of the $k$-space Hamiltonian are purely real,
%as the imaginary components are canceled out by symmetric hopping between lattice sites
which reflects the inherent symmetry constraints of the system. Breaking this symmetry enables asymmetric coupling, introducing imaginary parts into the off-diagonal terms of the $k$-space Hamiltonian. This approach significantly expands the model universality, allowing for more general physical scenarios.

\noindent\normalsize\textbf{Symmetrically-coupled lattice}
\normalsize
%频率维度的构成方法、MZI环

\begin{figure*}
    \centering
    \includegraphics[width=0.95
\textwidth]{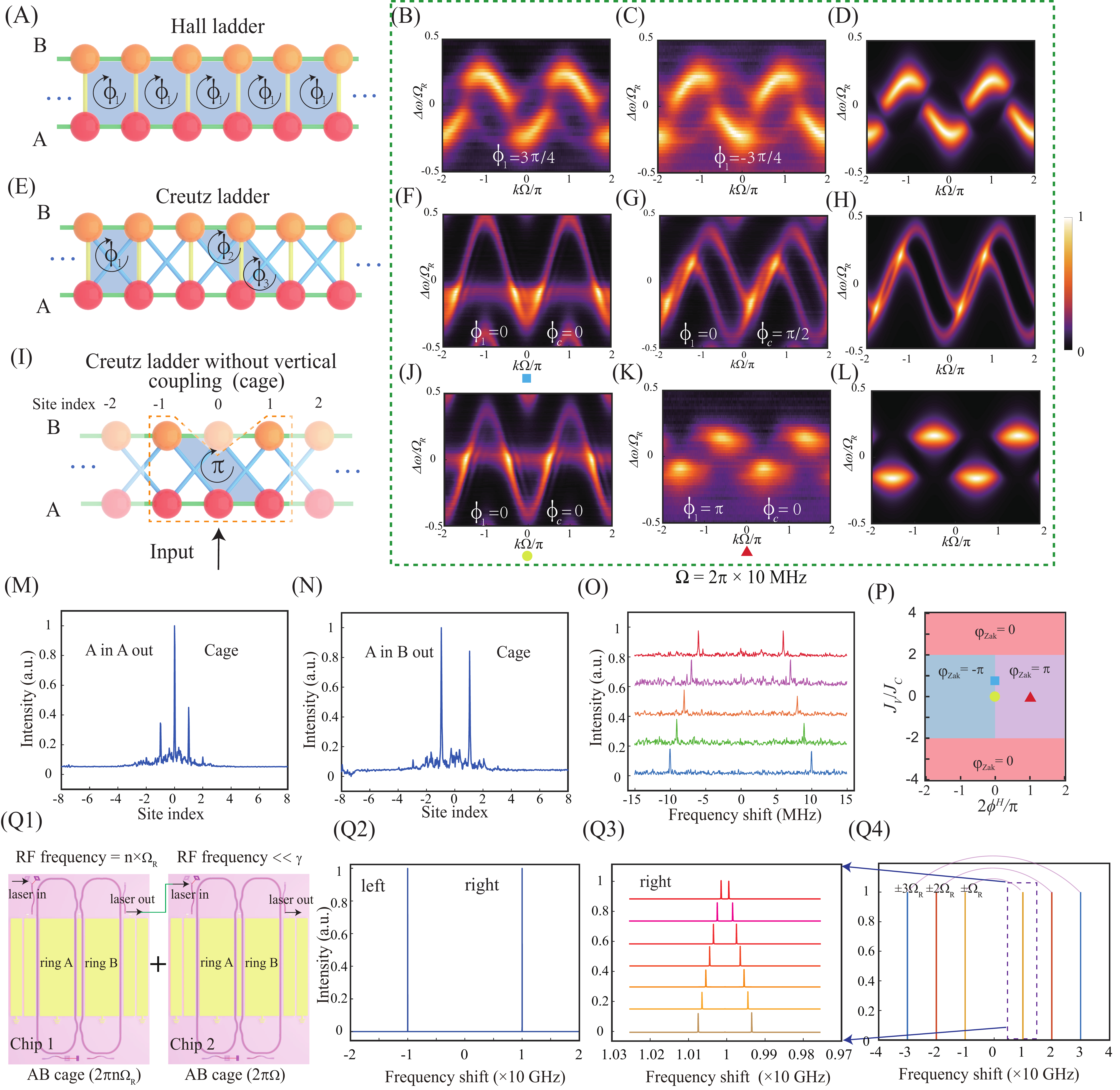}
    \caption{\textbf{Experimentally-obtained band structures of the Hall ladder and Creutz ladder, along with direct observation of the Aharonov-Bohm cage.}  (\textbf{A}) Illustration of the model of a Hall ladder. (\textbf{B, C}) The heat maps are the experimentally obtained band structures of the Hall ladder with coupling strength ratio $r=J^V/2J^H=0.9$, $J_H=0.085\Omega_R$, $\Omega$ = 2$\pi\times$10 MHz, and an effective magnetic flux $\phi_1$ being $-3/4\pi$ and $+3/4\pi$. (\textbf{E}) Schematic of Creutz ladder. The coupling types are of the same definition with those in Fig. \,\ref{fig:scheme} and $\phi_1$, $\phi_2$, $\phi_3$ are the potential gauges. (\textbf{F, G}) The heat maps display the general band structures of the Creutz ladder given different $\phi_A$, $\phi_B$ and $\phi_C$, where $J^V/2J^H=0.6$, $J^C/J^H=1$ and $J^H=0.11 \Omega_R$. (\textbf{I}) Illustration of the Aharonov-Bohm cage effect where the phase collected in a round trip in the blue area is $\pi$. When a probe light is input from Resonator A with frequency near the zero-index site, the distribution is caged within the dashed box ($0, \pm 1$ in sites A and $\pm 1$ in sites B). (\textbf{J, K}) The heat maps display the band structures of the Creutz ladder with $J^V=0$ and different $\phi_A$, $\phi_B$, $\phi_C$, where $J^C/J^H=1$ and $J^H=0.11 \Omega_R$. Notably, the flat band structure is measured when the magnetic flux, indicated by the blue shaded area in plot I, is $\pi$, where the Aharonov-Bohm cage effect is observed. (\textbf{D, H, L}) The theoretical results corresponding to the experimental data in their left neighboring panels. (\textbf{M, N}) The experimental readout of the site distributions from the drop ports of Resonators A and B measured by heterodyne detection. The results indicate a strong local effect of AB cage at the lower-index sites, and the higher-order modes are suppressed. The distribution results are plotted in the normalized linear coordinate. (\textbf{O}) Continuous sideband engineering with different $\Omega$ (2$\pi \times$ 6 $\sim$ 10 MHz) in AB cage (A in B out). (\textbf{P}) Distribution of Zak Phase in parameter space. The Zak phases corresponding to plots F, J and K are located at the positions indicated by the corresponding marks in this figure. (\textbf{Q1}) Schematic of piecewise-continuous optical frequency shifting enabled by cascaded devices. (\textbf{Q2}) Simulation of spectral output with frequency shift using only one MZI-assisted device via the inter-resonant mode under the AB cage configuration (RF1 = $\Omega_R$ = 2$\pi\times$ 10 GHz). (\textbf{Q3}) Simulation of spectral output with continuous frequency shift using two cascaded MZI-assisted devices (RF1 = $\Omega_R$ = 2$\pi \times$10 GHz, the positive branch, and RF2 = $\Omega = 2\pi \times$10 $\sim 70$ MHz). (\textbf{Q4}) Simulation of spectral output with segmented-continuous frequency shift using two MZI-assisted devices (RF1 = $\Omega_R = 2\pi\times 10,20,30$ GHz, RF2 = $\Omega = $ 2$\pi\times10\sim300$ MHz).}
    \label{fig:Hall}
\end{figure*}

Usually, frequency peaks spacing $\Omega_R$ (FSR) are precisely mapped to each site in the standard frequency lattice, and on resonance modulation, $\Omega = \Omega_R$ introduces the couplings (Fig. \,\ref{fig:scheme}A). However, the resonator is always lossy (with linewidth of $\gamma$), which broadens resonant peaks and can be utilized to relax constructive interference condition. This allows the resonator to support more frequency sites by putting the multiple coherent frequency sites in one broadened resonant peak. The discrete components with frequency $n\Omega$ $(n\in \mathbb{Z})$ can effectively simulate the behavior of discrete sites, which also enables the more feasible long-range couplings. As shown in Fig. \,\ref{fig:scheme}A, we combine the frequency domain and spatial domain. The driving RF signal of frequency $\Omega$, which is much smaller than the resonant linewidth, introduces couplings between the intra-resonant sites. We can depict the model with the horizontal, vertical, crossing coupling strengths ($J^H_{A}=J^H_{B}=J^H_{n\Omega}$, $J^V$, $J^{C}_{n\Omega}$), as well as the corresponding coupling phases ($\phi_{A(B),n\Omega}^H$, $\phi^V$, $\phi_{n\Omega}^{C}$), where $n$ denotes long-range coupling across $n$ sites. The vertical coupling arises from the spatial interaction between two resonators, the horizontal coupling is introduced by the RF modulation applied to the ring resonators, and the cross coupling is introduced by the RF modulation applied to the MZI. Here, the symmetry of the hopping remains, resulting in real-valued off-diagonal terms in the $k$-space Hamiltonian, i.e., $\phi^V=0$. Nearest-neighbor couplings are preserved along the horizontal and crossing directions. The $k$-space Hamiltonian of the Creutz ladder model is \cite{SM}:

\begin{equation}
\begin{aligned}
H_k =
-2\begin{bmatrix}
J^H\cos(k\Omega+\phi_A^H)& \frac{J^V}{2}+J^C\cos(k\Omega+\phi^C)\\
\frac{J^V}{2}+J^C\cos(k\Omega+\phi^C)& J^H\cos(k\Omega+\phi_B^H)
\end{bmatrix}.
\end{aligned}
\end{equation}
Ladders are among the simplest models while retaining important properties, providing a theoretical foundation and experimental pathway for exploring and understanding topological quantum states in low-dimensional systems.

%We denote the resonators by A and B, respectively. Here, $a_n (b_n)$ denotes the annihilation operator for the optical mode at frequency $n\Omega$ in Resonator A (B), with \textit{h.c.} representing the Hermitian conjugate. 

%The coupling strength in the Eq. (1) is respectively: $J^H_{A}(t)=J^H cos(\Omega t + \phi_{A})$, $J^H_{B}(t)=J^H cos(\Omega t + \phi_{B})$, $J^C(t)=J^C cos(\Omega t + \phi_{C})$.
%
\noindent\normalsize\textbf{Asymmetrically-coupled lattice}

\begin{figure*}[t]
    \centering
    \includegraphics[width=\textwidth]{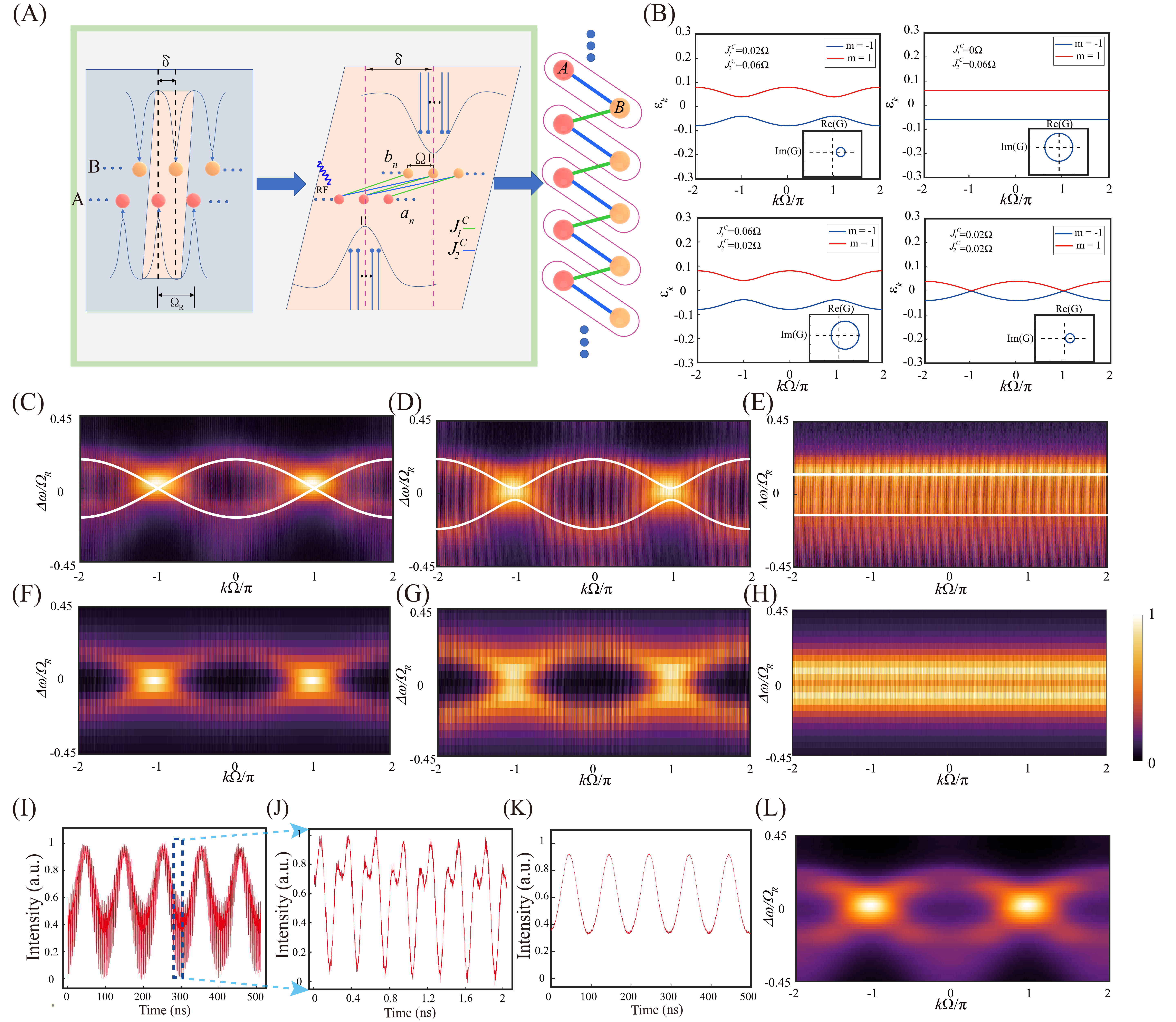}
    \caption{\textbf{Experimentally-obtained band structures of SSH model.} (\textbf{A}) Description of the optical frequency lattices in the periodically electro-optic modulated resonators. The resonant frequency peaks of Resonators A and B are detuned by $\delta$ using DC voltage. The construction of SSH model is realized using both intra-resonant and inter-resonant frequency sites, and among them, RF modulations at frequencies $\delta$ and $\delta + \Omega$ introduce two alternating types of couplings. The SSH model is shown in the right side of this panel, and the two sites circled in each purple box is a cell of this model. (\textbf{B}) The theoretical band structures (without noise) of the SSH model in plot \textbf{A} under different hopping strengths $J_1^C$ and $J_2^C$. Inset: the traces of (Re($G$),Im($G$)) evolving through the first Brillouin zone, where $G=\left|G\right|e^{i\varphi(k_f)}=J_1^{C}+J_2^{C}e^{ik_f\Omega+i\phi}$, and $\varphi(k_f)=\mathrm{arg}(G)$ is the argument of $G$ \cite{li2023light}. (\textbf{C-E}) Experimentally-observed band structure with various MZI modulation power. \textbf{C}: $J_1^{C}=J_2^{C}$, \textbf{D}: $J_1^{C}$ < $J_2^{C}$ and  \textbf{E}: $J_1^{C}=0$. The solid lines are the corresponding theoretical bands (without noise) solved from the $k$-space Hamiltonian. (\textbf{F-H}) The numerically-calculated results (with noise) for their corresponding upper panels. The horizontal axis stands for the quasimomentum and the vertical axis is the laser detuning. (\textbf{I,J}) The temporal transmittance signal waveform at different time scales, with the fixed laser detuning $\Delta \omega$. (\textbf{K}) The transmission spectra after applying a moving average to the data in Fig. \,\ref{fig:SSH}I. (\textbf{L}) The band structure diagram formed by stacking these moving-average transmission spectra. }
    \label{fig:SSH}
\end{figure*}

%Lattice structure formed by nonuniform connectivities between sites hold richer physics in real space. Most related studies have remained thoretical, including non-Hermitian SSH lattices and quadrupole higher-order topoloical insulators. In ref [], experimental work using supermode lattice points explored the extraction of the Zak phase in SSH, but did not directly observe or implement the band structure of SSH. 
To realize arbitrary couplings, we need to break the symmetry and introduce imaginary components into the off-diagonal terms of the $k$-space Hamiltonian. As an example, we directly simulate the SSH model and observe its band structures under the rotating wave approximation (RWA) \cite{SM}. The SSH model describes a one-dimensional asymmetrically-coupled lattice, where each unit cell contains two inequivalent atoms (A and B). As shown in Fig. \,\ref{fig:SSH}A, we first control the misalignment of the resonant peaks of Resonators A and B ($\delta$) using DC bias applied to both resonators. Then, by applying RF signals with frequency of $\delta$ to the MZI, coupling can be introduced between the detuned modes of the two resonators. When the applied RF modulation is a dual frequency signal $\Omega_1 =\delta$ and $\Omega_2 =\delta+\Omega$( $\Omega \ll \gamma$), under RWA, the SSH model in which sites are nonuniformly connected can be formed. Thus, we have demonstrated the construction of a model with asymmetric coupling. The $k$-space Hamiltonian of the constructed lattice is exactly the one of SSH model, in the following form:

\begin{equation}
\begin{aligned}
    &H_k = 
    \begin{bmatrix}
     0 & J_1^C + J_2^C e^{ik_f\Omega + i\phi}\\
    J_1^C  + J_2^C e^{-ik_f\Omega - i\phi} & 0
\end{bmatrix}.
    \label{eq.exp}
\end{aligned} 
\end{equation}
Building upon the aforementioned RF modulations, applying a modulation signal at frequency $\Omega$ to two resonators introduces long-range coupling into the SSH model, thereby can realizing the extended SSH (xSSH) model. We validate this through simulations provided in the Supplementary Materials \cite{SM}.

\begin{figure*}[t]
    \centering
    \includegraphics[width=\textwidth]{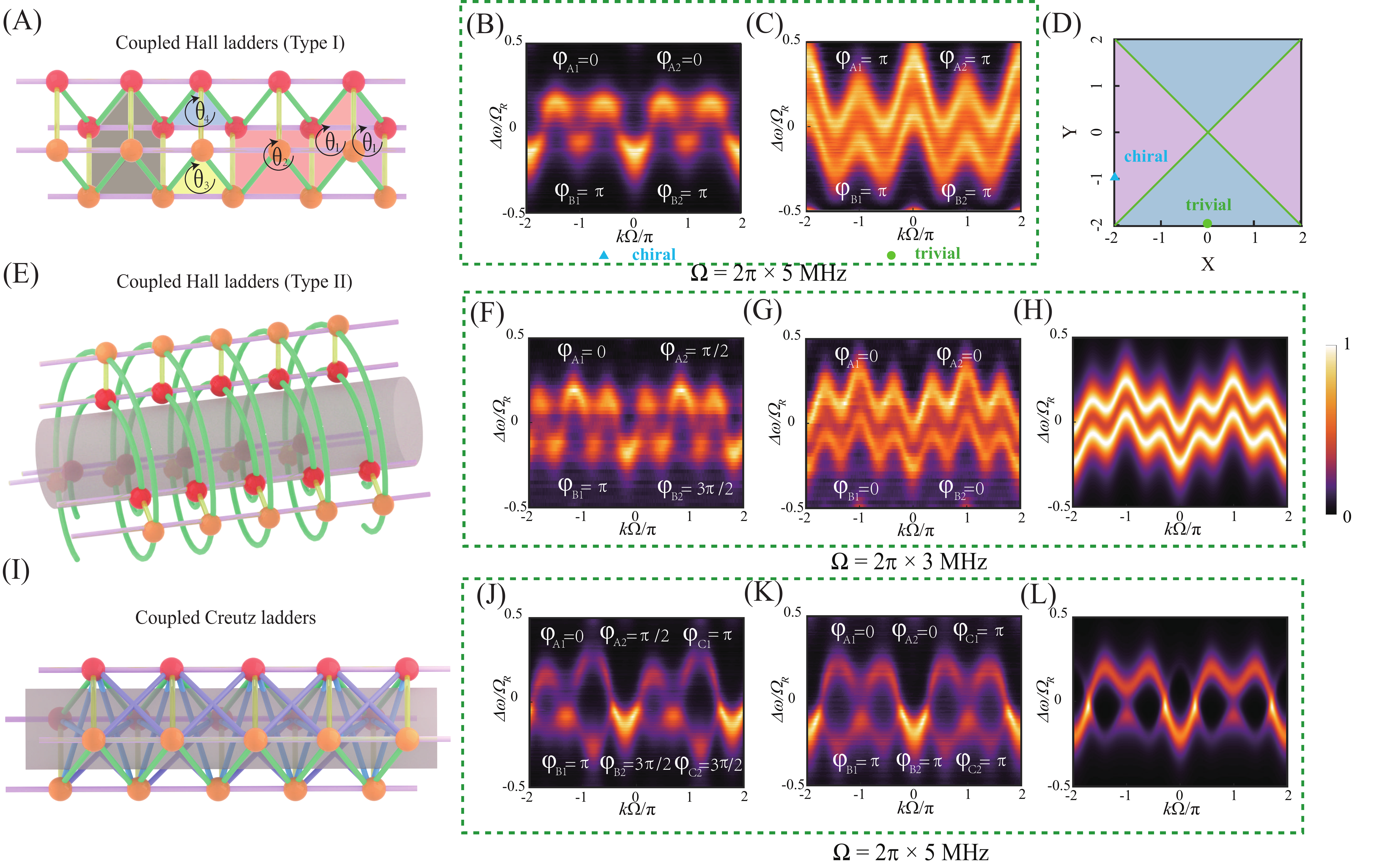}
    \caption{\textbf{Coupled Hall ladder, double-walled nanotube and coupled Creutz ladder.} (\textbf{A}) Illustration of the model of coupled Hall ladder. Four effective magnetic fluxes are defined in the coupled Hall ladder. $\theta_1$ and $\theta_2$ represent the effective magnetic fluxes of the Hall ladders with lattice constant of $\Omega$ and 2$\Omega$, respectively. $\theta_3$, $\theta_4$ represent the effective magnetic fluxes of the triangles within Resonators A and B.  (\textbf{B, C}) Experimental diagrams of the coupled-Hall-ladder-model band structures for different effective magnetic fluxes ($J_{A1}^H=0.14\Omega_R$, $J_{A2}^H/J_{A1}^H=1, J^V/2J_{A1}^H=0.7, J_{B1}^H=J_{A1}^H, J_{B2}^H=J_{A2}^H$). These results are corresponding to the band structures of the chiral phase and trivial phase in Fig. \,\ref{fig:longrange}D, respectively, with the same marks. (\textbf{D}) Phase diagram of the chiral edge current, which indicates versatile chiral edge states including chirality, chiral triviality, single-pseudospin enhancement (SPE) and complete suppression. $X = \rm{sgn}(\theta_2) - \rm{sgn}(\theta_1)$ and $Y = \rm{sgn}(\theta_3) - \rm{sgn}(\theta_4)$. (\textbf{E}) Double-walled nanotube lattice formed by the Hall ladder model with triple long-range coupling and nearest-neighbor coupling. (\textbf{F, G}) Experimental measurements of the influence of different effective magnetic fluxes in the double-walled nanotube lattice to its band structure ($J_{A1}^H=0.08\Omega_R$, $J_{A2}^H/J_{A1}^H=1.25, J^V/2J_{A1}^H=1.88, J_{B1}^H=J_{A1}^H, J_{B2}^H=J_{A2}^H$). (\textbf{H}) The theoretical result corresponding to the experimental data in the left-neighboring panel. (\textbf{I}) Complex lattice model formed by Creutz ladder with nearest-neighbor coupling and next-nearest-neighbor coupling. (\textbf{J, K}) The closing and opening of the band gap can be controlled by designing different effective magnetic flux of the lattice ($J_{A1}^H=0.05\Omega_R$, $J_{A2}^H/J_{A1}^H=1.3, J^V/2J_{A1}^H=1, J_{1}^C=J_{B1}^H=J_{A1}^H, J_{2}^C=J_{B2}^H=J_{A2}^H$ ). (\textbf{L}) The theoretical result corresponding to the experimental data on the left-neighboring panel.}
    \label{fig:longrange}
\end{figure*}

\noindent\large\textbf{Experiment and results}

\noindent\normalsize\textbf{Hybird synthetic-dimension chip}

\normalsize
First, we design and fabricate the double resonators coupled by an MZI with three sets of electrodes on the TFLN platform. The fabricated chip is shown in Fig. \,\ref{fig:scheme}B. The size of the entire chip is 10 mm $\times$ 20 mm, similar size to a coin. Fig. \,\ref{fig:scheme}C presents the optical microscope image of the racetrack microresonator. The illustration at the bottom of Fig. \,\ref{fig:scheme}C shows the images of the fabricated on-chip optical devices and electrodes, such as grating couplers, MZI modulation electrodes, etc. The experimental setup is illustrated in Fig. \,\ref{fig:scheme}D. DC and RF signals are simultaneously applied to the resonator and MZI through Bias-Tees. RF signals are generated by an arbitrary waveform generator (AWG) or microwave sources. The RF signal are amplitude-enhanced using power amplifiers before entering the Bias-Tees. The DC signals are used to align or misalign the center frequencies of the resonators, while the RF signals are used to establish couplings between the frequency-domain lattice sites. In the synthetic frequency sites, time is considered as the quasimomentum ($k$). With the scanning laser detuning $\Delta \omega$, the temporal data for each excited slice allows us to reconstruct the entire projected band structure \cite{Dutt2019,Dutt2020} through data stacking. The obtained pattern of this band structure is shown in the top-right Fig. \,\ref{fig:scheme}D. We use heterodyne detection to measure the optical-intensity distribution on the frequency lattice sites. An acousto-optic modulator (AOM) is used to shift the frequency of one portion of the input laser by 100 MHz, and it is interfered with the output light from the chip. This interfered light signal is then converted into electrical signal by a Photodetectors (PD) and recorded by an oscilloscope. The intensity distribution on frequency lattice can be obtained by Fourier transformation of this detected signal.

\noindent\normalsize\textbf{Symmetric coupling: Hall ladder and Creutz ladder}

Under the symmetric-coupling condition, the centers of resonant peaks of the two resonators are aligned. Experimentally, this alignment is achieved by adjusting DC signals applied to Resonators A and B, while the coupling strength between the two resonators is controlled via a DC bias on the MZI. By simultaneously applying RF modulation with frequency much smaller than the resonant-peak linewidth to Resonators A and B and adjusting the DC bias on the MZI to induce coupling between the resonators ($J^H_{A}(t)=J_A^H \cos(\Omega t + \phi_{A}^H)$, $J^H_{B}(t)=J_B^H \cos(\Omega t + \phi_{B}^H)$, $J^V(t)=J^V $, $J^C=0$, $\Omega = 2\pi \times 10 $ MHz $\ll \gamma$), the Hall ladder model is realized (Fig. \,\ref{fig:Hall}A). The two resonators can be alternatively depicted as a pseudospin. This enables us to observe various physical phenomena on the frequency lattice sites, such as spin-orbit coupling, equivalent magnetic flux, spin-momentum locking, etc. \cite{Dutt2019,Dutt2020}. Let $J^{H}=J^{H}_{A}=J^{H}_{B}$ with the proper choice of $\phi_1$. Fig. \,\ref{fig:Hall}B,C illustrate the projected band structures (to one of the pseudospin states, i.e., Resonator A) of the Hall ladder with various synthetic magnetic flux $\phi_1$ that indicates potential gauges. Spin-momentum locking is clearly seen in these experimental data (with $J^V/2J^H=0.9$). For instance, the transmission mode in Resonator A (representing one spin state) predominantly occupies the positive quasimomentum in the first Brillouin zone of the lower band, as seen in Fig. \,\ref{fig:Hall}B. %Additionally, we observe that the direction of spin-momentum locking switches for the upper band. (The probe laser selectively excites the band associated with pseudospin $n_+$, represented by the incident Resonator A (Fig. \,\ref{fig:Hall}E and F). In the first Brillouin zone, for $k>0$, the excitation occurs in the upper branch,  whereas for $k<0$, it occurs in the lower branch.  Upon inverting the magnetic flux to  $-\phi_1$, Resonator A turns into pseudospin $n_-$ due to symmetry.% 
When the synthetic magnetic flux is reversed, as shown in Figs. \,\ref{fig:Hall}B and \,C, the band structures in these two figures exhibit mirrored patterns to each other, indicating chiral spin-momentum locking. Fig. \,\ref{fig:Hall}D shows one of the theoretically calculated transmission spectra of the system (corresponding to Fig. \,\ref{fig:Hall}C) obtained via coupled-mode theory, demonstrating agreement with experimental data. 
\begin{comment}
The $k$-space Hamiltonian writes:
\begin{equation}
\begin{aligned}
H(k) = &-2J^H \left[\right. \cos(k\Omega)\cos(\phi_1/2)\,\mathbf{I}\\
&+\sin(k\Omega)\sin(\phi_1/2)\sigma_z \left.\right]-J^V\sigma_x,
    \label{eq.Hall}
\end{aligned}
\end{equation}
where $\sigma_x$ and $\sigma_z$ are Pauli matrices, $\mathbf{I}$ is the two dimensional identity matrix and $J^{H}=J^{H}_{A}=J^{H}_{B}$.
\end{comment}

Building upon the Hall ladder, the introduction of additional symmetric couplings, such as cross couplings, enable the construction of Creutz ladder. The Creutz lattice incorporates cross interactions between the legs, introduces two additional tunable magnetic flux parameters ($\phi_2$ and $\phi_3$), and exhibits distinct topological features, as illustrated in Fig. \,\ref{fig:Hall}E. The cross coupling strength is tuned by the amplitude of the RF signal applied to the MZI, while the magnetic flux is adjusted through the relative phase among the three driving signals. We again set equal coupling strength within each lattice. In Figs. \,\ref{fig:Hall}F and G, we activate all three types of couplings and alter magnetic flux to experimentally obtain band structures ($J^H_{A}(t)=J_A^H \cos(\Omega t + \phi_{A}^H)$, $J^H_{B}(t)=J_B^H \cos(\Omega t + \phi_{B}^H)$, $J^V(t)=J^V $, $J^C=J^C \cos(\Omega t + \phi^C)$, $J_A^H=J_B^H=J^H$, $J^V/2J^H=0.6$, $J^C/J^H=1$, $\Omega = 2\pi \times 10 $ MHz $\ll \gamma$). Aiming to configure the system to realize the topological flat band, it is essential to set $J^V=0$. By controlling the DC bias of MZI, we can realize zero vertical coupling between the two resonators, and the resulted lattice structure is shown in Fig. \,\ref{fig:Hall}I. Abundant band structures can also be observed, as shown in Fig. \,\ref{fig:Hall}J and K. In particular, we set $J^{C}=J^{H}$, $\phi^{C}=0$ and the phase collected in a round trip in the blue area is $\pi$ in Fig. \,\ref{fig:Hall}I. The topological flat band, shown in Fig. \,\ref{fig:Hall}K, exhibits energy independence of momentum, leading to zero group velocity. Similarly, Figs. \,\ref{fig:Hall}H and L display the theoretically calculated transmission spectra of the system, which are obtained via the coupled-mode theory, and show good agreement with the experimental data in the left neighboring panels. Moreover, the topological feature is also characterized in terms of AB cage effect. Under the same experimental configuration as that in Fig. \,\ref{fig:Hall}K and shown in Fig. \,\ref{fig:Hall}I, we inject a laser near the zeroth frequency mode into Resonator A, and simultaneously monitor the site distributions from  drop ports of both resonators via heterodyne detection. The results (Figs. \,\ref{fig:Hall}M, N) reveal the AB cage effect in synthetic intra-resonant frequency space, demonstrating clear suppression of intensity at the zeroth site of lattice B and higher-order sites of both lattices. In contrast, control experiments with zero magnetic flux ($\phi_A^{H}=\phi_B^{H}=0$) show delocalized site distributions (see Supplementary Materials \cite{SM}), confirming the absence of cage under non-topological conditions. The topological phase of the Creutz ladder is characterized by the Zak phase: $\varphi_{Zak}$ = 0 indicates a topologically trivial phase, while $\varphi_{Zak}$ = $\pm \pi$ corresponds to a non-trivial phase, as shown in Fig. \,\ref{fig:Hall}P in which three positions are marked. The band structure corresponding to Figs. \,\ref{fig:Hall}F and J are located on the phase transition line of $\varphi_{Zak}$ = $\pm \pi$ in the parameter space, whose positions are marked as the blue square and yellow circle, respectively with $J^V\neq 0$ and $J^V=0$. Their Zak phases are undetermined between the transition from $-\pi$ to $\pi$. The flat band, corresponding to Fig. \,\ref{fig:Hall}K and marked as the red triangle, has a Zak phase of $\pi$. These cases indicate the topologically non-trivial phase.

Due to the flexibility of the construction of intra-resonant frequency lattice, setting $\Omega$ within a certain range ($\Omega \ll \gamma$) can effectively define the lattice sites. With the AB cage effect of the Creutz ladder model, as shown in Fig. \,\ref{fig:Hall}O, adjusting $\Omega$ allows continuous sideband tuning. Experimentally, this tuning range of frequency is from 6 MHz to 10 MHz, but in principle it can be 300 MHz, as simulated in Supplementary Materials. Furthermore, we propose that cascading TFLN devices can enable piecewise-continuous optical frequency shifting. As shown in Fig. \,\ref{fig:Hall}Q1, we first configure the inter-resonant frequency lattice of Chip 1 to satisfy the topological flat-band conditions of the Creutz ladder model. Taking the optical mode distribution in Resonator B as the example, the optical frequency is shifted to $\pm n\Omega_R$ relative to the input frequency according to the applied RF signal ($n\Omega_R$). The output light from Resonator B of Chip 1 (as shown in Fig. \,\ref{fig:Hall}Q2, here we set $n=1$ as the example) is then used to input Chip 2, which is also configured to meet the topological flat band condition in its intra-resonant lattice. Around the two frequency peaks input to Resonator B ($\pm \Omega_R$), two new sets of frequency modes with spacing $\Omega$ are generated, as shown in Fig. \,\ref{fig:Hall}Q3 (take the $+\Omega_R$ branch as example). By combining the RF signals applied to control Chip 1 and Chip 2, the output optical spectrum can be tuned both largely and (local) precisely, as shown in Fig. \,\ref{fig:Hall}Q4. Using this proposal, the application of segmented continuous optical frequency shifting can be achieved \cite{li2025weakly,yang2025}.

\noindent\normalsize\textbf{Symmetric Coupling: Long-range coupling}

Next, we simulate the band structure of lattice models with long-range coupling. This clearly demonstrates the benefits of an intra-resonant lattice incorporating low-frequency modulation on integrated chips. By introducing the next-nearest-neighbor coupling into the Hall ladder, we can construct a pseudo-three-dimensional frequency lattice \cite{ye2025observing,liu2025}. To achieve long-range coupling, we apply the RF signal with multiharmonic components expressed as $J^H_{A}(t)=J^H_{A1} \mathrm{cos}( \Omega_1 t + \varphi_{A1})+J^H_{A2} \mathrm{cos}( \Omega_2 t+\varphi_{A2})$, $J^H_{B}(t)=J^H_{B1} \mathrm{cos}( \Omega_1 t+\varphi_{B1})+J^H_{B2} \mathrm{cos}( \Omega_2 t+\varphi_{B2})$, $J^C(t)=0$, and for the coupled Hall ladder model $\Omega_2 = 2\Omega_1=2\pi \times 10$ MHz. Compared to previous works, the frequency requirement for RF modulation is reduced by about three orders of magnitude. The lattice structure is depicted in Fig. \,\ref{fig:longrange}A. These fluxes are related to the modulation signals by $\theta_1=\varphi_{B1}-\varphi_{A1}$, $\theta_2=\varphi_{A2}-\varphi_{B2}$, $\theta_3=\varphi_{B2}-2\varphi_{B1}$ and $\theta_4=2\varphi_{A1}-\varphi_{A2}$. By configuring these magnetic fluxes in the sub-ladders, the states of the coupled Hall ladder can be conveniently manipulated, and its band structure is shown in Figs. \,\ref{fig:longrange}B, C. These diverse chiral edge states \cite{liu2025} are summarized in a phase diagram in Fig. \,\ref{fig:longrange}D. By introducing third-nearest-neighbor couplings ($\Omega_2 = 3\Omega_1=2\pi \times 9$ MHz) into the Hall ladder, one can construct a double-walled nanotube lattice, with the corresponding lattice model illustrated in Fig. \,\ref{fig:longrange}E. Subsequent controlled modulation of RF signals enables systematic engineering of effective magnetic flux parameters, manifesting in the characteristic band structure evolution of the double-walled nanotube lattice, as mapped in Figs. \,\ref{fig:longrange}F, G. For the intra-resonant frequency-lattice sites, we can also straightforwardly introduce long-range couplings into the Creutz ladder via modulation signals $J^H_{A}(t)=J^H_{A1} \mathrm{cos}( \Omega_1 t+\varphi_{A1})+J^H_{A2} \mathrm{cos}( \Omega_2 t+\varphi_{A2})$, $J^H_{B}(t)=J^H_{B1} \mathrm{cos}( \Omega_1 t+\varphi_{B1})+J^H_{B2} \mathrm{cos}( \Omega_2 t+\varphi_{B2})$, $J^C(t)=J^C_{1} \mathrm{cos}( \Omega_1 t+\varphi_{C1})+J^C_{2} \mathrm{cos}( \Omega_2 t+\varphi_{C2})$ with $\Omega_2 = 2\Omega_1= 2\pi \times 10$ MHz. By modulating effective magnetic flux parameters, we can obtain the band structure under different effective flux in the coupled Creutz ladder, as shown in Figs. \,\ref{fig:longrange}J, K. Based on the corresponding $k$-space Hamiltonian and experimental parameters, the theoretical band structures shown in Figs. \,\ref{fig:longrange}H and L exhibit excellent consistency with the experimental results in the left neigboring panels.

\begin{comment}
the Hamiltonian in $k$-space writes:
\begin{equation}
\begin{aligned}
H(k) = &-2J^H \left[\right. \cos(k\Omega)\cos\frac{\phi_1}{2}\,\mathbf{I}+\sin(k\Omega)\sin\frac{\phi_1}{2}\sigma_z \left.\right]\\
&-\left[\right. J^V+2J^C\cos(k\Omega+\phi_2) \left.\right]\sigma_x,
    \label{eq.Creutz}
\end{aligned}
\end{equation}
where $\phi_2=\phi^{D}-\phi_A^{H}$, $\phi_3=\phi_B^{H}-\phi^D$ and we have set $\phi_2=-\phi_3$.  
\end{comment}

%The main reasons of the deviation may be the imperfection of the MZI, the slight misalignment of the resonators and the leakage caused by high order transitions of the phase modulation.

\noindent\normalsize\textbf{Asymmetric coupling: Su-Schrieffer-Heeger model}

To break the symmetry and establish asymmetric coupling, we intentionally misalign the central frequencies of the two resonators, by shifting them with $\delta$ ($\gamma$ < $\delta$ < $\Omega_R$/2). As a result, the vertical coupling $J^V$ disappears, and the cross coupling $J^C$ arises from dual-frequency RF modulation on the MZI with frequencies $\delta$ ($J_1^C$) and $\delta+\Omega$ ($J_2^C$). This modulation drives the unidirectional hoppings on the lattice sites, which is formed by the small frequency separation $\Omega$ ($\Omega$ $\ll$ $\gamma$) inside the resonant peaks of both Resonators A and B. For example, the $J_2^C$ coupling only enables the site $a_n$ at frequency $n\Omega$ to hop to the neighboring resonator site $b_{n+1}$ at $\delta+(n+1)\Omega$ (forward hopping), but $J_1^C$ only enables $a_n$ to hop to $b_{n}$ at $\delta+n\Omega$ (backward hopping). Consequently, forward and backward hoppings are separated and can be tuned independently, thus breaking the symmetry. %Particularly, by adjusting the DC bias settings on the Resonators A and B, the central frequencies of the resonators are detuned by $\delta$ ($\delta$ > $\gamma$). As noted, applying RF signals to the MZI induces cross coupling. Introducing dual RF modulation with a small frequency offset $\Omega$ ($\Omega$ $\ll$ $\gamma$ ) between the arms enables precise synthesis of
This forms the SSH lattice configurations, as systematically mapped in the right-side of Fig. \,\ref{fig:SSH}A, in which the two sites circled in one purple box form a cell of the SSH model. Equation (S17) reveals a two-band system (m $=\pm$1) in Fig. \,\ref{fig:SSH}B (see Supplementary Materials). The complex parameter G, which defines the off-diagonal elements of the Hamiltonian's $k$-space representation, traces circular orbits in the Re(G)-Im(G) space during Brillouin zone traversal, with radius $J_2^{C}$ and centered at ($J_1^{C}$,0). The topological character of these orbits is determined by the inclusion or exclusion of the zero point, i.e., (0,0). Encircling the zero point ($J_1^{C}$ < $J_2^{C}$) yields winding number W = 1 (non-trivial), while the exclusion of this point ($J_1^{C}$ > $J_2^{C}$) gives W = 0 (trivial). Notably, the exchange of $J_1^{C}\leftrightarrow J_2^{C}$ leads to a topological phase transition between trivial and non-trivial states, despite the band structure remains unchanged, as demonstrated in Fig. \,\ref{fig:SSH}B. 

Through the DC tuning protocol, we implement $\delta=2 \pi \times3$ GHz inter-resonator detuning between the resonant peaks of Resonators A and B, and the lattice constant is set as $\Omega= 2 \pi \times 10$ MHz. The RF excitation parameters are configured as $J^H_{A}(t)=0$, $J^H_{B}(t)=0$, $J^C(t)=J^C_{1} \mathrm{cos}( \Omega_1 t)+J^C_{2} \mathrm{cos}( \Omega_2 t+\phi_C)$, $\Omega_1 = \delta= 2\pi \times 3$ GHz, $\Omega_2=\delta+\Omega= 2\pi \times 3.01$ GHz. In our engineered SSH-model framework, the first Brillouin zone in $k$-space is defined by \( k\Omega/\pi \in [-1,1] \).  By configuring the oscilloscope monitoring window to $2\pi$/$\Omega$ timescale, we obtain sliced transmission spectra of the SSH-model band structure. Experimentally, the intra-cell and inter-cell hopping strengths in the SSH model are tuned by controlling the RF powers $P_{\Omega_1}$ and $P_{\Omega_2}$ applied to the MZI. As shown in Figs. \,\ref{fig:SSH}C-E, when $J_1^{C}$ = $J_2^{C}$, the band gap closes. As $J_2^{C}$ increases beyond $J_1^{C}$, the gap begins to open. When $J_1^{C}$ decreases to 0, the SSH model reduces to isolated site pairs with flat bands. The solid lines in the figure represent theoretical curves that agree well with the experimental data. Figs. \,\ref{fig:SSH}F-H present the corresponding numerically calculated results. We have mentioned that the band structure is stacked through the transmission spectra at the timescale of $2\pi/\Omega$ by scanning laser. In each slice (e.g., Fig. \,\ref{fig:SSH}I, the same condition as Fig. \,\ref{fig:SSH}C), although high-frequency ($\delta$) oscillation noise exists (see Supplementary Materials), it does not affect band structure identification. Zooming into Fig. \,\ref{fig:SSH}I reveals finer spectral lines and subwaveforms (Fig. \,\ref{fig:SSH}J), arising from the high-frequency oscillation ($\delta$). The high-frequency noise can be removed using a moving average method. Fig. \,\ref{fig:SSH}K displays the transmission spectra after applying a moving average to the data in Fig. \,\ref{fig:SSH}I. The corresponding band structure diagram formed by stacking these moving average transmission spectra is presented in Fig. \,\ref{fig:SSH}L.

Under the RWA, we achieve the SSH model by combining intra-resonant with inter-resonant site couplings. The corresponding band structure was directly observed via time-resolved transmission spectra. In our system, establishing long-range coupling is straightforward. Building upon the aforementioned RF modulation, applying the RF signal at frequency $\Omega$ to the resonators enables more complex couplings between different cells, thereby realizing the extended SSH (xSSH) model. We have validated the effectiveness of this approach through simulations, as shown in the Supplementary Materials.

\noindent\large\textbf{Discussions}

\normalsize
%总结
We have extended the intra-resonant frequency lattice to higher dimensions (combined with spatial domain and inter-resonant frequency mode) in this experiment. By employing the MZI-assisted resonators and RF modulation frequency significantly below their linewidth, we simulate abundant lattice models including Hall ladder, Creutz ladder, and observe topological flat bands with AB cage effect. Furthermore, we propose that the flexibility of the intra-resonant frequency lattice enables sideband engineering application via the AB cage effect, allowing frequency tuning of synthetic sidebands. Then, we propose that cascading our chips enables the application of segmented continuous optical frequency shifting, which is demonstrated through simulations. This showcases the potential of our synthetic frequency-dimension simulator in spectral-engineering control applications. Similarly, by utilizing intra-resonant frequency lattice, we can readily introduce long-range coupling experimentally. We simulated lattice models with long-range coupling based on the Hall ladder and Creutz ladder frameworks, such as those resembling the double-walled nanotubes. Finally, we have combined intra-resonant and inter-resonant frequency sites for the first time. By programming our device, we have constructed nonuniform-connectivity lattices and demonstrated simulations of the SSH and xSSH model lattices. Our method enables the implementation of complex lattice models, including symmetric, asymmetric and long-range couplings. This provides a typical example for building large-scale on-chip synthetic frequency-dimension quantum simulators.

%展望

\noindent\large\textbf{Methods}

\normalsize
Our photonic device is constructed using an x-cut thin-film lithium niobate on insulator (TFLN) substrate. This platform consists of a 400 nm thick lithium niobate layer atop a 4.7 $\rm{\mu}$m buried $\text{SiO}_2$ layer, which itself is supported by a 525 $\rm{\mu}$m silicon base. The resonator structures are defined through electron-beam lithography (EBL), using hydrogen silsesquioxane (HSQ) as a negative-tone resist. After patterning, approximately 200 nm of the lithium niobate layer is selectively removed by argon-based inductively coupled plasma reactive ion etching (ICP-RIE). To finalize the structure, any remaining HSQ resist and redeposited byproducts are eliminated using a wet chemical etching step. %To reduce the length of bonding wires, we need to extend the eletrodes to the edges of the chip, which requires some of the electrodes to cross waveguides. To address this, we build ``bridges'' above some waveguides.
The device is coated with a bilayer of ultraviolet photoresists, LOR5A and S1813, through spin coating. Using markless lithography, ``bridges'' structures are defined to allow electrode crossing over the optical waveguides. A 750 nm thick $\text{SiO}_2$ layer is then deposited and patterned via a lift-off technique. Following this, copper electrodes with a thickness of 800 nm are formed using a comparable process. These electrodes are subsequently connected to a high-frequency printed circuit board (PCB) through wire bonding.

Tunable microwave sources and an arbitrary waveform generator are employed to produce the RF signals, which are subsequently amplified using high-gain amplifiers. The laser is output by a distributed feedback (DFB) laser, after passing through a polarization controller, and is coupled into the device via a grating coupler. For low-frequency signal measurements, the output is either directly detected or first processed through a heterodyne detection setup before being recorded using a photodiode and an oscilloscope. In contrast, high-frequency signals are captured and analyzed using a high-speed sampling oscilloscope.

\textbf{Acknowledgements} 

This work is supported by the Innovation Program for Quantum Science and Technology (No. 2021ZD0301200), the National Natural Science Foundation of China (Nos. 12174370, 12174376, 11821404, 12304546, and 124B2082), the Youth Innovation Promotion Association of Chinese Academy of Sciences (No. 2017492), Anhui Provincial Natural Science Foundation (No. 2308085QA28), China Postdoctoral Science Foundation (No. 2023M733412). Quantum Science Center of Guangdong-Hong Kong-Macao Greater Bay Area Research Initiation Fund (No. QD2305001) and Guangdong Provincial Quantum Science Strategic Initiative (No. GDZX2403003). This work was partially carried out at the USTC Center for Micro and Nanoscale Research and Fabrication.

\bibliography{reference}
\end{document}